\documentstyle[12pt]{article}
\begin{document}

\overfullrule 0 mm
\language 0
\centerline {{ \bf THE LORENTZ-DIRAC EQUATION:}}
\centerline {{ \bf THE INSTABILITY OF "PHYSICAL" SOLUTION}}
\vskip 0.3 cm

\centerline {{  Alexander A. Vlasov}}
\centerline {{  High Energy and Quantum Theory}}
\centerline {{  Department of Physics}}
\centerline {{  Moscow State University}}
\centerline {{  Moscow, 119899}}
\centerline {{  Russia}}
\vskip 0.3 cm
{\it The instability of "physical" preaccelerative solution  of the 
Lorentz-Dirac equation is explicitly shown }

03.50.De

Since the famous Dirac's paper on relativistic radiation reaction in
classical electrodynamics, many textbooks and research articles were
published on that theme. Among them are [1-11], where one can find 
the discussion of the related problems: mass renormalization, 
runaway solutions and the use of the advanced interaction.

However one problem, to my opinion, draws not enough attention in 
the literature - this is the problem of instability of 
preaccelerative "physical" solution  of relativistic Lorentz-Dirac 
equation.

In this article on the example of one-dimensional motion of a 
particle the instability of such solutions will be explicitly shown.

Let the point particle with mass $m$ and charge $e$ move under 
some external force $F$ along $x$-axis. The relativistic 
Lorentz-Dirac  equation reads: 
$$ {m \dot v \over (1- (v/c)^2)^{3/2} }= F
+{2e^2 \over 3 c^3}\left[ {\ddot v \over (1- (v/c)^2)^{2}} +
{3v (\dot v)^2 \over c(1- (v/c)^2)^{3}}  \right] \eqno (1) $$
here dot means differentiation with respect to $t, \ v=\dot x$.

Take the dimensionless variables $\eta$ and $\tau$,  dimensionless 
force $f$ and dimensionless radiation parameter $\gamma$ and 
 introduce scale multiplies $a,b,p$:  
 $$a \sim x_{cl}, \ b \sim x_{cl}, \ x_{cl}={e^2\over mc^2}, \ 
 p={a\over b}$$ $$x=a\eta, \ ct=b\tau, \  f=F{b^2 \over amc^2}, \ 
 \gamma={2\over 3b}x_{cl}, $$
here $x_{cl}$ - the classical particle radius.

In terms of new variables equation (1) becomes 
  $$ { \ddot \eta \over (1- (p\dot\eta)^2)^{3/2} }= f 
+\gamma\left[ { \dot {\ddot \eta }\over (1- (p\dot\eta)^2)^{2}} +
{3\dot \eta (\ddot \eta)^2 \over (1- (p\dot\eta)^2)^3}  \right] \eqno
(2) $$
here dot means differentiation with respect to   $\tau$.

With the help of relativistic "velocity" $u$:
  $$u ={ \dot \eta \over (1- 
(p\dot\eta)^2)^{1/2} } ={ v/(cp)\over (1- 
(v/c)^2)^{1/2} } \eqno(3)  $$ 
 "acceleration" $w$:  
$$w={du \over d\tau} \eqno(4)$$ and dimensionless proper time $s$:  
 $${ b \over c(1- (v/c)^2)^{1/2} } {d \over 
dt}=(1+ (p u)^2)^{1/2} {d \over d\tau}= {d\over ds} $$
equation (2) can be put in more simple form:
$$w=f +\gamma {dw \over ds} \eqno (5)$$ 
The "solution" of (5) is obvious:
  $$w=w(s)=-{1\over \gamma}\exp{(s/\gamma)} 
\int\limits_{s_0}^{s}dxf(x)\exp{(-x/\gamma)} \eqno(6)$$
with "initial" value $w_0$:
$$w_0=\left({ \ddot \eta \over (1- (p\dot\eta)^2)^{3/2} 
}\right)_{s=0}=
{1\over \gamma}\int\limits_{0}^{s_0}dxf(x)\exp{(-x/\gamma)} $$

Integration of (4), taking into consideration  (3) and  (6), 
yields the following "solution" (strictly  speaking, the integral 
equation) for particle velocity $v$:  $${1\over p}\ln { \sqrt{ 
 {(1+v/c)(1-v_{0}/c) \over (1-v/c)(1+v_{0}/c)} } }=$$ 
$$\int\limits_{0}^{s}dzf(z) 
-\exp{(s/\gamma)}\int\limits_{0}^{s}dzf(z)\exp{(-z/\gamma)}
+{1\over \gamma}w_0\left(\exp{(s/\gamma)}-1\right) \eqno(7) $$
here $v_0$- the "initial" velocity.

The form  of Lorentz-Dirac equation  similar to (7) was found in 
Parrott's works [4-7]. The form (7) is convenient for 
 analysis. Thus from it immediately follows that:

(i) The peculiar features of Lorentz-Dirac equation do not 
qualitatively depend on scale multiplies $a,\ \ b,\ \ p$,  so these 
features are valid as for "small", so for "large", "classical" 
distances.

(ii) If  conditions of a problem permit to consider the limit $s\to 
\infty$ (all integrals in (7) are not divergent for $s<\infty$ ) 
then all "solutions" (7) must be "runaway" - $|v|\to c$, with one 
exception.

(iii)  This exception is the particular case of zero asymptotic 
value of "acceleration"  $w$:  $s_0= \infty$ in (6) and
$$w_0={1\over \gamma} 
\int\limits_{0}^{\infty}dxf(x)\exp{(-x/\gamma)} \eqno (8)$$ 
With (8), the R.H.S. of (7) takes the form
 $$\int\limits_{0}^{s}dzf(z) 
+\exp{(s/\gamma)}\int\limits_{s}^{\infty}dzf(z)\exp{(-z/\gamma)}
-{1\over \gamma}w_0 \eqno(9) $$
 Then for $s\to \infty$ and for "well defined" force $f$ the 
 velocity $v$ does not reach $c$. But the price for this is
the preacceleration and backward in time 
integration with accompanying paradoxes (some of 
them are discussed in [7,8,9]).

(iv) In literature the "solution" (9) is often called  
"physical", but from equation (7) it is easy to see that (9) is 
unstable under small deviations of "acceleration" $w$ from zero 
value at infinite "future": due to (7) these initially small at 
$s=+\infty$ deviations $\delta$ grow at least as $e^{\delta}$.

The instability of the "physical" solution one can also verify with 
the help of numerical calculation: the instability was found in 
[10] and in recent calculations, made by author for the force $f$ 
formed by external charge at rest.

Following (iv) one can state that there are no stable "nonrunaway" 
solutions of Lorentz-Dirac equation, at least in one-dimensional 
case.

Further, if one rejects unstable solutions and 
takes only "runaway" 
solutions, then for $s\to \infty$   (if problem under consideration 
permits this) these solutions:

(i) may give infinite particle "acceleration" $w$ at "future" (and 
many examples of physical force $f$ do lead to this);

(ii) tend to light cone with $|v| \to c$ (and as particle 
trajectory tends to light cone, the retarded time $t_{ret}$ near 
particle trajectory may differ strongly from the laboratory time 
$t$).

These features can contradict the standard methods to obtain 
relativistic Lorentz-Dirac equation, when the following 
conditions must be fulfilled (see, for ex., [11]):  
$$t_{ret}-t=\epsilon,\ \ \epsilon \to 0,\ \ 1 \gg 
w\epsilon/(1-(v/c)^2)^{1/2} $$

Then, if it is so, two possibilities are left:

(i) to consider unstable solutions as corresponding to the real 
physics;

(ii) to reject Lorentz-Dirac equation and to search for new one.

    \vskip 3 mm
  \begin{enumerate}
  \item
  F.Rohrlich, {\it Classical Charged Particles}, Addison-Wesley,
  Reading, Mass., 1965.
\item
A.Sokolov, I.Ternov, {\it Syncrotron Radiation}, Pergamon Press,
    NY, 1968. A.Sokolov, I.Ternov, {\it Relativistic
Electron} (in russian), Nauka, Moscow, 1983.
\item D.Ivanenko, A.Sokolov,  {\it Classical field theory} (in
russian), GITTL, Moscow, 1949
 \item S.Parrott, {\it
Relativistic Electrodynamics and Differential Geometry},
 Springer-Verlag, 1987.

 \item C.Teitelboim, Phys.Rev., 1970, D1, p.1572; D2, p.1763.
\item  E.Glass, J.Huschilt and G.Szamosi, Am.J.Phys., 1984, v.52, 
p.445. 
 \item S.Parrott, Found.Phys., 1993, v.23, p.1093.  
\item 
W.Troost et al.,  preprint hep-th/9602066, 1996 
\item A.Vlasov, 
preprint hep-th/9702177, 1997 
\item J.Kasher, Phys.Rev.,1976, D14, 
p.939.  
\item S. de Groot, L.Suttorp {\it Foundations of 
Electrodynamics}, North-Holland, Amsterdam, 1972 

\end{enumerate}

 \end{document}